# The nature and origin of time-asymmetric spacetime structures[*]

H. D. Zeh (University of Heidelberg)

www.zeh-hd.de

**Abstract:** Time-asymmetric spacetime structures, in particular those representing black holes and the expansion of the universe, are intimately related to other arrows of time, such as the second law and the retardation of radiation. The nature of the quantum arrow, often attributed to a collapse of the wave function, is essential, in particular, for understanding the much discussed "black hole information loss paradox". However, this paradox assumes a new form and might not even occur in a consistent causal treatment that would prevent the formation of horizons and singularities.

A "master arrow", which combines all arrows of time, does not have to be identified with the direction of a formal time parameter that serves to define the dynamics as a succession of global states (a trajectory in configuration or Hilbert space). It may even change direction with respect to a fundamental *physical* clock, such as the cosmic expansion parameter if this was formally extended either into a future contraction era or to negative "pre-big-bang" values.

## 1 Introduction

Since gravity is attractive, most gravitational phenomena are asymmetric in time: objects fall down or contract under the influence of gravity. In General Relativity, this asymmetry leads to drastically asymmetric spacetime structures, such as future horizons and future singularities as properties of black holes. However, since the relativistic and nonrelativistic laws of gravitation are symmetric under time reversal, all time asymmetries must arise as consequences of specific (only seemingly "normal") *initial* conditions, for example a situation of rest that can be prepared by means of other ar-rows of time, such as friction. Otherwise, conclusions like gravitational contraction would have to apply in both directions of time. Indeed, the symmetry of the gravitational laws does allow objects to be thrown up, where their free motion could in principle *end* by another external intervention, or the conceivable existence of "white holes", which would have to contain past singularities and past horizons.

---

[*] arXiv:1012.4708v11. V5 was published in the *Springer Handbook of Spacetime Physics* (A. Ashtekar and V. Petkov, edts. – Springer 2014). See also the "Note added after publication" at the end of this text!

The absence of such past horizons and singularities from our observed universe (except, perhaps, for a very specific big bang singularity) must be regarded as a time asymmetry characterizing our global spacetime (see Sects. 2 and 4), while Einstein's field equations would not only admit the opposite situation (for example, inhomogeneous past singularities), but also many solutions with mixed or undefined arrows of time – including closed time-like curves and non-orientable spacetimes. Therefore, the mere possibility of posing an "initial" condition is exceptional in general relativity from a general point of view. I will here not discuss such mathematically conceivable solutions that do not seem to be realized in Nature, but instead concentrate on models that come close to our universe – in particular those which are globally of Friedmann type. A specific arrow characterizing a Friedmann universe is given by its expansion (unless this would be reversed at some time of maximum extension – see Sect. 4).

In many cases, non-gravitational arrows of time remain relevant for the evolution of gravitating bodies even *after* the latter have been prepared in an appropriate initial state. This applies, in particular, to strongly gravitating objects, such as stars, whose evolution is essentially controlled by thermodynamics (emission of heat radiation into the cold universe). The relation between the electrodynamic and thermodynamic arrows (retardation and the second law, respectively)[1] is quite obvious in this case.

Gravitating systems are nonetheless thermodynamically unusual in possessing negative specific heat.[2] This means, for example, that stars become hotter when losing energy by emitting heat, and that satellites accelerate as a consequence of friction in the earth's atmosphere. It can best be understood by means of the virial theorem, which states in its nonrelativistic form, and for forces that decrease with distance according to the inverse square law (that is, gravitational and Coulomb forces), that all bound states have to obey the relation $\overline{E_{pot}} = -2\overline{E_{kin}}$, where the overbar means averaging over (quasi) periods of time. Therefore,

$$E = E_{pot}(t) + E_{kin}(t) = \overline{E_{pot}} + \overline{E_{kin}} = \frac{1}{2}\overline{E_{pot}} = -\overline{E_{kin}} \propto -T \quad (1)$$

When losing thermal energy by radiation, these systems must gain twice as much from gravitational contraction in order to maintain a quasi-stable state. Nonrelativistically, this negative heat capacity could be bounded by means of other (repulsive) forces that become relevant at high densities, or by the Pauli principle, which controls the density of electrons in white dwarf stars or solid planets, for example. Relativistically, even these limits will break down at a certain mass, since (1) relativistic degeneracy must ultimately lead to the creation of other particles, while (2) the potential energy of repulsive forces will itself gravitate, and for a sufficiently large mass overcompensate any repulsion. Therefore, it is the thermodynamic arrow underlying thermal radiation that requires evolution of gravitating systems towards the formation of black holes. Classically, black holes would thus define the final states in the evolution of gravitating systems.

## 2 Black Hole Spacetimes

The metric of a spherically symmetric vacuum solution for non-zero mass is shown in Fig. 1 in Kruskal coordinates $u$ and $v$. This diagram represents the completed Schwarzschild metric in the form

$$ds^2 = \frac{32M^2}{r}e^{-r/2M}\left(-dv^2 + du^2\right) + r^2\left(d\theta^2 + \sin^2\theta d\phi^2\right), \quad (2)$$

where the new coordinates $u$ and $v$ are in the external region ($r > 2M$) related to conventional Scharzschild coordinates $r$ and $t$ by

$$u = e^{r/4M}\sqrt{\frac{r}{2M} - 1}\cosh\left(\frac{t}{4M}\right) \quad (3a)$$

$$v = e^{r/4M}\sqrt{\frac{r}{2M} - 1}\sinh\left(\frac{t}{4M}\right). \quad (3b)$$

Each point in the diagram represents a sphere with surface $4\pi r^2$. Note that $r$ and $t$ interchange their roles as space and time coordinates for $r < 2M$, where $2M$ is the Schwarzschild radius. All parameters are given in Planck units $h/2\pi = G = c = 1$.

As Nature seems to provide specific initial conditions in our universe, it may thereby exclude all past singularities, and hence all past event horizons. This initial condition would immediately eliminate the Schwarzschild-Kruskal vacuum solution that is shown in the Figure, but we may instead consider the future evolution of a spherically symmetric mass distribution initially at rest, such as a dust cloud. It would classically collapse freely into a black hole, as quantitatively described by the Oppenheimer-Snyder scenario[3] (see left part of Fig. 2). The vacuum solution (2) is then valid only outside the surface of the dust cloud, but this surface must according to a classical description fall through the arising horizon at some finite proper time, and a bit later hit the future singularity.

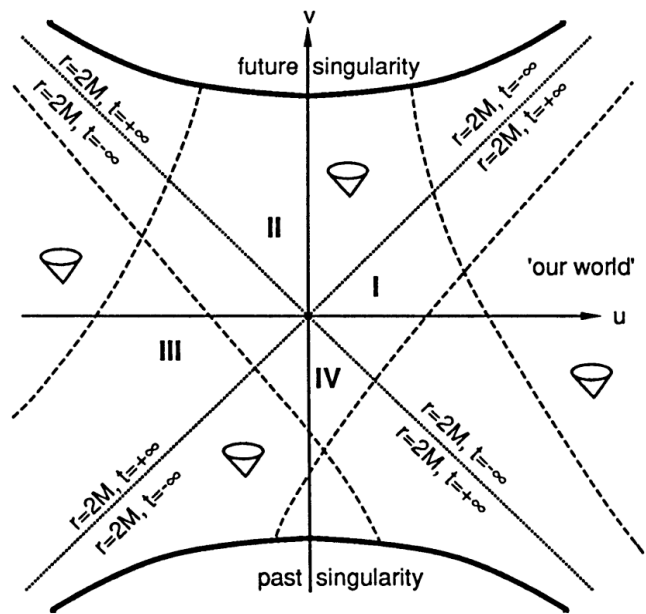

**Fig. 1:** Complete formal continuation of the Schwarzschild solution by means of unique Kruskal coordinates. Quadrants *I* and *II* represent external and internal parts, respectively, of a classical black hole. *III* is another asymptotically flat region, while *IV* would describe the interior of a "white hole". In this diagram, fixed Schwarzschild coordinates $r$ and $t$ are represented by hyperbola and straight lines through the origin, respectively. World lines of local objects could start at $t = -\infty$ in *I* or at $t = +\infty$ in *III*, or at $r = 0$ on the past singularity in *IV*, while they must end at $t = +\infty$ or $-\infty$ in *I* or *III*, respectively, or at a second singularity with coordinate value $r = 0$ in *II*. On time-like or light-like curves intersecting one of the horizons at the Schwarzschild radius $r = 2M$, the value of the coordinate $t$ jumps from $+\infty$ to $-\infty$ at the rim of quadrant *I*, or from $-\infty$ to $+\infty$ at the rim of quadrant *III*, where $t$ decreases in the global time direction.

For a cloud of interacting gas molecules, this gravitational collapse would be thermodynamically delayed by the arising pressure, as indicated in the Introduction. Gravitational radiation would lead to the loss of any kind of macroscopic structure, while whatever remains would become unobservable to an external observer. Although thermodynamic phenomena control the loss of energy by radiation during most of the time, the asymmetric absence of *past* singularities represents a fundamental cosmological initial condition. However, a conceivable white hole initiated by a past singularity that *completely* represented a time-reversed black hole would even require anti-thermodymics and coherently incoming advanced radiation. One may suspect that all these various arrows are related to one another, thus defining a common "master arrow".

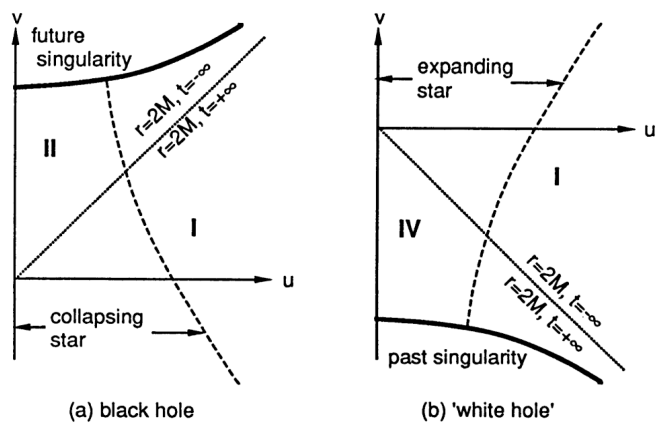

**Fig. 2:** Oppenheimer-Snyder type spacetimes of a black and a "white" hole.

Since it would require infinite Schwarzschild coordinate time for an object to reach the horizon, any message it may send to the external world shortly before it does so would not only be extremely redshifted, but also dramatically delayed. The message could reach a distant observer only at increasingly later stages of the universe. (An apparatus falling into a galactic size black hole could even send messages for a considerable length of *proper* time before it would approach the horizon.) So all objects falling into the black hole must effectively disappear from the view of mortal external observers and their descendants, even though these objects never seem to reach the horizon according to their rapidly weakening, but in principle still arriving signals. The only asymptotically observable properties of the black hole are conserved ones that have early enough caused effects on the asymptotic metric or other asymptotic fields, namely angular momentum and electric charge. This time-asymmetric consequence is known as the "no-hair theorem" for black holes. During cosmological times, a black hole accumulating ion-ized interstellar matter may even lose its charge and angular momentum, too, for statistical and dynamical reasons.[4] Only its mass and its center of mass motion would then remain observationally meaningful. A black hole is usually characterized by its center of mass motion and its long-lasting properties, namely its mass *M*, charge *Q*, and angular momentum *J*, in which case its "Kerr-Newman metric" is explicitly known. The internal topological structures of these metrics for *J ≠ 0* and/or *Q ≠ 0* are radically different from that of the Kruskal geometry in Fig. 1, thus raising first doubts in the validity of these classical continuations inside the horizon.

It is important, though, to keep in mind the essential causal structure of a black hole: its interior spacetime region *II* never enters the past of any external observer, that is, it will never become a "fact" for him. While the whole exterior region *r > 2M* can be completely foliated by means of "very nice" space-like slices according to increasing Schwarzschild or similar time coordinates with -∞ < *t* < +∞, the interior can then be regarded as its *global future* continuation beyond the event horizon, where increasing time can be labeled by the Schwarzschild coordinate *r* decreasing from *r = 2M* to *r = 0.* This structure must be essential for all causal considerations that include black holes – not least for their own fate (Sect. 3). In the classical scenario, the internal state of a black hole would be completely determined by the infalling matter, which could even depend on our "free" decisions about what to drop into a black hole. Nonetheless, properties of this infalling matter would then irreversibly become "irrelevant" to all external observers – a term that is also used to define a generalized concept of coarse graining required for the concept of physical entropy in statistical thermodynamics.[5]

## 3 Thermodynamics and the Fate of Black Holes

In the classical picture described above, a black hole would represent a perfect absorber at zero temperature. This picture had to be corrected when Bekenstein and Hawking demonstrated,[6] the latter by explicitly taking into account quantum fields other than gravity, that a black holes must possess finite temperature *T* and entropy *S* proportional to its surface gravity *κ* and surface area *A*, respectively:

$$T = \frac{\hbar \kappa}{2\pi k_B} \to \frac{\hbar c^3}{8\pi G k_B} \frac{1}{M} \quad , \tag{4a}$$

$$S = \frac{k_B c^3 A}{4\hbar G} \rightarrow \frac{4\pi k_B G}{\hbar c} M^2 \quad . \tag{4b}$$

Here, $\kappa$ and $A$ are known functions of $M$, $Q$ and $J$, while the explicit expressions given on the right hand side of the arrow hold for Schwarzschild black holes ($Q = J = 0$) and with respect to spatial infinity (that is, by taking into account the gravitational redshift). This means, in particular, that a black hole must emit thermal radiation (Hawking radiation) proportional to $T^4 A$ according to Stefan-Boltzmann's law, and therefore, that it lives for a very large but limited time of the order $10^{65}(M/M_{sun})^3$ years. For stars or galaxies this is very many orders of magnitude more than the present age of the universe of about $10^{10}$ years, but far less than any Poincaré recurrence times for such macroscopic systems. So one has to be careful about what is meant by "asymptotic" in different contexts.

Even these large evaporation times will begin to "count" only after the black hole has for a very long time to come *grown* in mass by further accreting matter[7] (including anti-matter if it becomes available during the black hole´s very long journey through the universe) – at least until the cosmic background temperature has dropped below the very small black hole temperature. Although evaporation times are thus extremely long, all radiation registered by an external observer must have been caused outside the horizon. Schwarzschild times represent proper times of distant observers in the rest frame of the black hole, but the spacelike slices that they define may be consistently continued inwards while remaining outside the horizon in order to form a complete foliation of the whole external region *I.* By definition, they would then all have to include the center of the collapsing matter at a pre-horizon stage. However, a horizon and its interior region *II* could never form if the black hole's energy was indeed radiated away *before* any infalling matter arrived at the classically predicted horizon in the sense of this global dynamical foliation. Although such matter may need only seconds of proper time to reach the classically expected horizon, there must always exist simultaneities which include events on the late pre-horizon part of its trajectories as well as external ones in the far future – including those at $t \approx 10^{65}$ years or more from now. This singular gravitational time dilation does *not* require any extreme spacetime curvature in the region where it applies. Attempts to find forces or stress terms that prevent infalling matter from crossing the horizon for this purpose would be reminiscent of Poincaré's search for forces to explain the Lorentz contraction. So what happens to matter that seems to fall into the black hole (and that may even be entangled with matter that remains outside)?

Schwarzschild simultaneities may thus be counterintuitive. One may also use time translation invariance of the external region of the Kruskal type diagram (Figs. 1 or 2a) in order to define the time coordinate *v = t = 0* to coincide with an external time close to the peak of the Hawking radiation (in the very distant future from our point of view) without coming any closer to the horizon that is defined by the remaining black hole mass. Assuming that one can neglect any quantum uncertainty of the metric (which must in principle arise in quantum gravity), all infalling matter that had survived the radiation process so far would at this coordinate time *v = 0* be in the very close vicinity of the center. Therefore, this simultaneity represents quite different proper times for the various parts of infalling matter even for a collapsing homogeneous dust cloud – and even more so for later infalling things. Proper times are irrelevant for the global geometrodynamics. Most of the black hole's original mass-energy must already exist in the form of outgoing Hawking radiation at this coordinate time, and may even have passed any realistic "asymptotic" observer. In order to be observed by him, it must have its causal root outside of an horizon.

Black hole radiation is again based on the radiation arrow of retardation, but its conventional formulation also depends on a quantum arrow that is defined by the statistical interpretation of quantum mechanics. A pure quantum state forming a black hole would accordingly decay into many fragments (mainly photons), described by a statistical ensemble of different emission times – similar to the ensemble of all potential outcomes in a series of measurements, or to the cooling of a highly excited quantum state by means of many stochastic radiation events.[8] However, an *apparent* ensemble is defined according to an appropriate concept of coarse graining already for an outgoing *pure* state that is the result of a unitary description (without any pair creation *events* that would also cause "real" ingoing negative energy radiation). In quantum theory, one usually neglects in this sense (that is, one regards as irrelevant for the future) the entanglement between decay fragments. Such a coarse-graining (neglect of information) does not only formally justify the concept of growing "physical" entropy in spite of the conservation of a pure global state,[5] but also the phenomenon of decoherence (which has here to occur in an external photon detector, for example). In contrast to the global ensemble entropy that is conserved under unitary dynamics (and vanishes for a pure state), physical entropy is defined as an extensive quantity that gives rise to the local concept of an entropy density which neglects information about correlations – just as Boltzmann's *µ*-space distribution does. The thermal Hawking radiation can thus not represent a *proper* mixture for the same reason why decoherence does not explain a "real" collapse of the wave function.[39] The major difference between the decay of highly excited states of normal matter and the evaporation of black holes is that the latter's unitary dynamics is not explicitly known (and occasionally even questioned to apply).

The thus described situation is nonetheless much discussed as an "information loss paradox for black holes".[9] Its consequences are particularly dramatic if one *presumes* the existence of a black hole interior region that *would* necessarily arise in the absence of Hawking radiation; matter (and the "information" it may represent) *could* then not causally escape any more. This questionable presumption (based on classical singularity theorems) is often tacitly introduced by using "nice slices" that are defined to avoid the singularity but may, in contrast to our "very nice slices", intersect the thus also presumed horizon. Unitary description means, however, that the information which defines the initial pure state is mostly transformed into non-local entanglement. Global unitarity thus leads to a *superposition* of "many worlds" which thereafter remain dynamically autonomous, and which may include different versions of the "same" observers – thus *physically* justifying decoherence as describing an apparent collapse.[39] The replacement of this superposition by an ensemble of many *possible* worlds according to a fundamental statistical interpretation (a real collapse of the wave function) would instead objectively annihilate the information contained in their relative phases, and in this way introduce a *fundamental* (law-like) dynamical time asymmetry. Recall that the Oppenheimer-Snyder model, on which the nice slices are based, precisely neglects the energy loss of the black hole by Hawking radiation. Although the ("back") reaction of the metric in response to radiation loss may in principle require quantum gravity, my argument about the non-formation of a horizon is here only based on the local conservation of momentum-energy in a situation where this may not have to be questioned.

Instead of assuming an external vacuum when calculating probabilities for Hawking radiation, one should take into account the presence of infalling matter, in which case some kind of internal conversion might lead to its annihilation. (The conservation of baryon number etc. would have to modify the Hawking radiation, and may thus lead to an essentially different scenario.) A similar scenario has recently been postulated as a novel kind of physics close to the horizon (called a "firewall").[10] While this firewall was meant to prevent an observer from remaining intact when falling in, it should according to my earlier proposal (see earlier versions of this paper, available at arxiv:1012.4708v1 or v2) convert *all* infalling matter into outgoing radiation. Note that the *local* Bekenstein-Hawking temperature diverges close to the horizon, and would therefore describe all kinds of particle-antiparticle pairs in a non-inertial frame (such as at a fixed distance). As long as some internal conversion of this kind cannot be excluded, there is no reason to speculate about black hole remnants, superluminal tunneling, or a *fundamental* violation of unitarity that would go beyond decoherence (that is, beyond a mere dislocalization or "globalization" of superpositions that just renders them irrelevant for local observers).[11] Unitarity can only apply to the global "bird's perspective" that includes all Everett branches, and it cannot lead to any kind of "double-entanglement".[12]

What might remain as a "remnant" according to this semi-classical description of black hole evolution on very nice slices is a *massless pointlike* curvature singularity, since the Riemann tensor of the Schwarzschild metric is proportional to $M/r^3$, and hence diverges for $r=2M \rightarrow 0$. This singularity signals a break-down of the semi-classical description of geometrodynamics at this final stage only. For example, quantum gravity would require a boundary condition for the timeless Wheeler-DeWitt wave function, which cannot distinguish between past and future singularities (see Sects. 4 and 5). This might lead to an effective final condition that affects black holes "from inside" in an anticausal manner.[13] Any inwards-directed (hence virtual) negative energy radiation compensating the emission of Hawking radiation according to some pictures could then "recohere" the effective black hole state in order to lower its entropy in accordance with both the mass loss and Bekenstein's relation (4b).

Note that the concept of an S-matrix would also be unrealistic for macroscopic objects, such as black holes. Because of their never-ending essential interaction with their environments, they can never become asymptotically isolated (the reason for their ongoing, locally non-unitary decoherence). The extreme lifetime of black holes means that the information loss problem is clearly an academic one: any apparently lost information would remain irrelevant for far more than $10^{65}$ years, and it could hardly ever be exploited even if it finally came out as entangled radiation. It can only describe *one* superposition of "many worlds" which form an apparent ensemble. The "Page time",[14] when the entanglement between the residual black hole and its emitted radiation is assumed to be maximal, can therefore not have any consequences for the *observed* black hole.

Several physicists (including myself) used to see a problem in the equivalence principle, which requires that observers or detectors freely falling into the black hole do *not* register any Hawking radiation. Some even concluded that the mass-loss of black holes, too, must then be observer-dependent (not very appropriately called "black hole complementarity"). However, this conclusion appears to be wrong. While the equivalence between a black hole and a uniformly accelerated detector (as regards their specific radiation) must indeed apply to the local laws, it can in general *not* apply to their boundary conditions. An observer or detector fixed at some distance from the black hole would not be immersed in *isotropic* heat radiation, since this radiation is coming from the direction of the black hole surface, which would cover most of the sky only for an observer very close to the horizon. Even though the freely falling detector may then not register any radiation, the latter's effect on *fixed* detectors, or its flux through a fixed sphere around the black hole, must exist objectively – just as the clicks of an accelerated detector in an inertial vacuum (attributed to Unruh radiation) could be noticed by *all* observers, regardless of their own acceleration. They all have to agree that the energy absorbed by the accelerated detector must be provided by the rocket engine and, analogously, that the Hawking net flux of energy requires an observer-independent mass loss of the black hole. Therefore, the dynamically resulting spacetime geometry (including consequences of stochastic measurement outcomes) is also objectively defined. The freely falling observer would furthermore hear the clicks of fixed detectors occurring at a very fast rate, and so as being caused by a very intense outward flux according to his proper time. For the same reason, matter at the outer rim of a collapsing dust cloud can at late Schwarzschild times not experience any gravitational field, as there is practically no gravitating energy left inside its present position any more. Hence, it can never cross a horizon.

In this way, the phenomenon of black holes from the point of view of external observers is *consistent* with the fate of a freely falling observer, who may either soon in his proper time have to be affected himself by the internal conversion process, or otherwise have to experience the black hole surface very rapidly shrinking – finally giving rise to extreme tidal forces – and disappearing before the observer's remains arrive. Note that the auxiliary concept of an event horizon changing in time is in principle ill-defined, since a horizon is already a spacetime concept. The apparent black hole surface *r=2M(u)*, where *M(u)≈M(t)* characterizes the corresponding Vaidya metric, while *u* is here the outgoing Eddington-Finkelstein coordinate, may nonetheless shrink adiabatically in order to dis-appear before any infalling matter has got a chance to enter the region *r≤2M(t)* for any finite coordinate time *t*.

If the freely falling observer could survive the internal conversion process, he would have travelled far into the cosmic future in a short proper time because of the quasi-singular time dilation. On the other hand, no theory that is compatible with the equivalence principle can describe baryon number non-conservation in the absence of a singularity. Because of the huge life time of black holes this problem may perhaps be solved in connection with that of the matter-antimatter asymmetry in our universe. All symmetries may in principle be broken by the effective non-unitarity characterizing the dynamics of *individual* Everett branches. This last remark might also be relevant for the above mentioned possibility of anti-causality (recoherence) required by an apparent future condition that is in accord with a timeless Wheeler-DeWitt equation (see Sect. 5); recoherence would require a re-combination of different Everett worlds.

Roger Penrose had compared black hole entropy numerically with that of matter in the universe under normal conditions.[15] Since the former is according to (4b) proportional to the square of the black hole mass, macroscopic black hole formation leads to a tremendous increase of physical entropy. As thermodynamic entropy is proportional to the particle number, it is dominated in the universe by photons from the primordial cosmic radiation (whose number exceeds baryon number by a factor *$10^9$*). If our observable part of the universe of about *$10^{79}$* baryons consisted completely of solar mass black holes, it would possess an entropy of order *$10^{98}$* (in units of $k_B^{-1}$), that is, *$10^{10}$* times as much as the present matter entropy that is represented by *$10^{88}$* photons. Combining all black holes into a single one would even raise this number to *$10^{121}$*, the highest conceivable entropy for this (perhaps partial) universe unless its volume increased tremendously.[4,7,16] If entropy is indeed a measure of probability, any approximately homogenous matter distribution would be extremely improbable except for densities much lower than at present (at a very late stage of an eternally expanding universe). Therefore, the homogeneity of the initial universe is usually regarded as the "fundamental improbable initial condition" that explains the global master arrow of time if statistical reasoning is applicable to the future (see Sect. 4). However, its relationship to the thermodynamically important condition of absent or "dynamically irrelevant" non-local initial correlations (or entanglement in the quantum case) seems to be not yet fully understood. If the two entropy concepts (black hole and thermodynamic) are to be compati-ble, the entropy of the final (thermal) radiation must be greater than that of the black hole, while the latter has to exceed that of any kind of collapsing and infalling matter.

**4 Expansion of the Universe**

The expansion of the universe is a time-asymmetric process, but in contrast to most other arrows it forms an *individual* phenomenon rather than a whole class of similar observable ones, such as black holes, radiation emitters, or steam engines. It may even change its direction at some time of maximum extension, although present astronomical observations may indicate that the expansion will last forever. A homogeneous and isotropic Friedmann universe is in classical GR described by the dynamics of the expansion parameter $a(t)$ in accordance with the time-symmetric "energy theorem" for $\ln[a(t)]$,

$$(da/adt)^2/2 = (4\pi/3)\rho(a)+\Lambda/6-k/2a^2 \quad , \tag{5}$$

where $\rho$ is the energy density of matter, $\Lambda$ the cosmological constant, and $k=0,\pm1$ the sign of the spatial curvature. The value of the formal "total energy" (the difference of both sides of the equation) is thus fixed and vanishes in general-relativistic cosmology. Penrose's entropy estimates then demonstrate that the homogeneity assumed in Eq. (5) is extremely improbable from a statistical point of view. Therefore, it must be unstable under the influence of gravity (in spite of being dynamically consistent).

In accordance with a homogeneous initial matter distribution, Penrose postulated that free gravitational fields vanished exactly at the Big Bang. These free fields are described by the *Weyl tensor*, that is, the trace-free part of the curvature tensor. The trace itself (the Ricci tensor) is locally fixed by the stress-energy tensor of matter according to the Einstein field equations. The Weyl tensor, on the other hand, is analogous to the divergence-free part of the electrodynamic field tensor $F^{\mu\nu}$, since the divergence $\partial_\mu F^{\mu\nu}$ (the trace of the tensor of its derivatives) is similarly fixed by the charge current $j^\nu$. Therefore, the *Weyl tensor hypothesis* is analogous to the requirement of an absence of any initial electromagnetic radiation, a condition that would allow only the retarded electromagnetic fields of all sources in the universe to exist. This universal retardation of radiation had indeed been proposed *as a law* by Planck (in a dispute with Boltzmann),[17] and later by Ritz (in a dispute with Einstein),[18] in an attempt to *derive* the thermodynamic arrow. However, Boltzmann and Einstein turned out to be right, since the retarda-tion can in turn be understood as a *consequence* of the presence of thermodynamic absorbers.[1] In cosmology, this includes the absorber formed by the radiation era, which would not allow us to discover any conceivable earlier electromagnetic radiation. In contrast, the early universe seems to be transparent to *gravitational* radiation, including that which might have been created in the Big Bang.

Note that the low entropy and the corresponding homogeneity of the universe can *not* be explained by an early cosmic inflation era (as has occasionally been claimed) if this inflation was deterministic and would thus have conserved ensemble entropy.

Although our universe may expand forever, the idea of its later recontraction is at least conceptually interesting. Thomas Gold first argued that the low entropy condition at high density should not be based on an absolute direction of time, and hence be valid at a conceivable Big Crunch as well.[19] The latter would then be observed as another Big Bang by observers living during the formal contraction era if the Weyl tensor was required to vanish there as well. Gold's scenario would not only require a thermodynamic transition era without any well-defined arrow in our distant future – it would also pose serious consistency problems (similar to Wheeler and Feynman's absorber theory[1]), since the extremely small initial probability for the state of the universe would have to be squared if the two conditions are statistically independent of one another.[20] If nonetheless true, it would have important consequences for the fate of matter falling into massive black holes. If such black holes survived the mentioned thermodynamic transition era at the time of maximum extension because of their long evaporation times (cf. Sect. 3), they would according to the global dynamics enter an era with reversed arrows of time. However, because of the transparence of the late universe to light, they would "receive" coherent advanced radiation from their formal future even before that happens. This advanced radiation must then "retro-cause" such massive black holes to expand again in order to approach a state of homogeneity in accordance with the final condition.[21] In mathematical terms, their horizon is not "absolute" in this case even in the absence of any black hole evaporation.

A reversal of the arrow of time may not only occur in the distant future, but may also have occurred in the past. Several *pre-big-bang* scenarios have been discussed in novel and as yet speculative theories. Usually, one thereby identifies the direction of the formal time parameter with the direction of the physical arrow of time. For example, ac-cording to arguments first used in loop quantum cosmology,[22] the configuration space for Friedmann type universes may be doubled by interpreting formally negative values of the cosmic expansion parameter *a* as representing negative volume measures. The cosmic dynamics can then be continued backwards in time beyond the Big Bang into its mirror image by "turning space inside out" (turning right-handed triads into left-handed ones) while going through *a = 0* even in a classical picture. For this purpose, the classical dynamical description (5) would have to be modified close to the otherwise arising singularity at *a = 0* – as it is indeed suggested by loop quantum gravity. However, if the "initial" conditions responsible for the arrow of time are assumed to apply at the situation of vanishing spatial volume, the arrow would formally change direction, and |*a*| rather than *a* would represent a physical cosmic clock. Observers on both temporal sides of the Big Bang could only remember events in the direction towards *a = 0*. Another possibility to avoid the singularity is a repulsive force acting at small values of *a*,[23] which would lead to a Big Bounce with similar conceivable consequences for the arrow of time as the above model that involves space inversion.

In cosmology, quantum aspects of the arrow of time must again play an important role. According to the Copenhagen interpretation, there is no quantum world – so no complete and consistent cosmic history would be defined any more when quantum properties become essential. In other orthodox interpretations, the unitary evolution of the quantum state is repeatedly interrupted by measurements and similar time-asymmetric events, when the wave function is assumed to "collapse" indeterministically. The consequences of such stochastic events on quantum cosmology would be enormous, but as long as no collapse mechanism for the wave function has been confirmed, one has again arrived at an impasse. Going forward in time may be conceptually simple in such asymmetric theories, since one just has to "throw away" all components of the wave function which represent the not "actualized" *potential* outcomes, while going backwards would require all these lost components to recombine and dynamically form local superpositions again. So one has at least to keep them in the cosmic bookkeeping – regardless of whether they are called "real" (as in the Everett interpretation) or not. Going back to the Big Bang by means of the unitary dynamics would require *all* those many "worlds" that have ever been thrown away in the orthodox description during the past of our universe, while one would have to throw away others when formally going backwards beyond the Big Bang in order to obtain an individual quasi-classical "pre-big-bang history".

In other words, a unitary continuation beyond the Big Bang can only describe the complete Everett superposition of worlds on both sides of the Big Bang, but hardly any individually observed quasi-classical worlds. The corresponding master arrow of time would thus not only affect all realms of physics – it must be truly universal in a much deeper sense: it can only have "multiversal" meaning. The same multiversality was required in a unitary black hole evolution of Sect. 3, and it does, in fact, apply to the unitary quantum description of all macroscopic objects, when irreversible decoherence mimics a collapse of the wave function and thereby explains classicality.

The time direction of Everett's branching of the wave function that is based on decoherence requires a homogeneous initial quantum state (presumably at *a = 0*), which does not contain any nonlocal entanglement that might later have local effects. Quantum dynamics will then lead to decoherence (the in practice irreversible dislocalization of superpositions), and thereby "intrinsically" break various global symmetries – possibly even in the form of many different quasi-classical "landscapes", which can only represent different branches of *one* symmetric superposition.

**5 Quantum Gravity**

General Relativity has traditionally been considered in a block universe picture, but because of the hyperbolic type of Einstein's field equations it is a dynamical theory just as any other field theory. Its explicit dynamical description, which requires a non-Lorentz-invariant *form*, was completed by Arnowitt, Deser and Misner (ADM).[24] This Hamiltonian formulation is a prerequisite for the canonical quantization of the theory. I shall here regard the result of this quantization procedure as an *effective* quantum theory, without discussing any attempts of a justification in terms of theories that may possibly be exact but have no empirical support as yet (such as string theory or loop quantum gravity).

The ADM formalism is based on an arbitrary time-like foliation of spacetime that has to be chosen "on the flight", that is, *while* solving an initial value problem numerically. (A similar freedom was used in Sect. 3 for the choice of very nice slices.) If the dynamics of matter is also defined, this construction must lead to a unique (foliation-independent) spacetime geometry, while the spatial metric on the *chosen* space-like slices represents the corresponding dynamical variables. The latter can be described by a symmetric ma-trix $h_{kl}(x_m)$ – with $k,l,m$ running from *1* to *3*. Three of its six independent matrix elements represent the choice of unphysical coordinates, two would in the linear approximation correspond to the spin components of a gravitational wave (±2 with respect to the direction of propagation for a plane wave), while the remaining one can be regarded as a measure of "many-fingered" physical time (metric distance between adjacent space-like slices). The corresponding canonical momenta $\pi^{kl}$ define the embedding of the spatial metric into spacetime and the arbitrary propagation of spatial coordinates. The dynamics can then be formulated by means of the Hamiltonian equations with respect to an arbitrary time parameter *t* that formally distinguishes different slices in a given foliation. These Hamiltonian equations are equivalent to Einstein's field equations. In contrast to metric time, the parameter *t* is geometrically or physically meaningless, and can therefore be replaced by any monotonic function *t' = f(t)* – including its inversion.

Note that when Special Relativity is said to abandon the concept of absolute time, this statement refers only to the concept of absolute simultaneity, while proper times, which control all motion according to the principle of relativity, are still assumed to be given "absolutely" by the fixed Lorentz metric. This remaining absoluteness is thus abandoned only in General Relativity, where the metric itself becomes a dynamical object like matter, as described by the ADM formalism. The absence of an absolute time parameter (here represented by its reparametrizability) was already required by Ernst Mach. Julian Barbour, who studied its consequences in much historical detail,[25] called it "timelessness". However, a complete absence of time would remove any possibility to define an arrow, while a one-dimensional (dynamical) succession of states, characterized by an arbitrary parameter, still allows one to define a time direction asymmetry.

The invariance of the theory under spatial coordinate transformations and time reparametrization is warranted by four constraints for the matrix $h_{kl}(t)$, called momentum and Hamiltonian constraints, respectively. They may be regarded as initial conditions, but they are conserved in time. In particular, the Hamiltonian constraint assumes the form

$$H(h_{kl},\pi_{kl}) = 0 \quad . \tag{6}$$

When quantized,[26] and when also taking into account matter variables, this constraint translates into the Wheeler-DeWitt equation,

$$H \, \Psi(h_{kl}, matter) = 0 \quad , \tag{7}$$

which means that the time-dependent Schrödinger equation becomes trivial,

$$\partial \Psi/\partial t = 0 \quad . \tag{8}$$

Even the time parameter *t* has now disappeared, because there are no parametrizable trajectories representing cosmic histories any more in quantum gravity. Only this drastic property, which is a *quantum* consequence of classical reparametrizability, can be regarded as a formal "timelessness".

The timelessness of the Wheeler-DeWitt wave function has been known at least since 1967, but it seems to have originally been regarded as "just formal". A time parameter was often smuggled in again in various ways – for example in terms of parametrizable Feynman paths, by means of semiclassical approximations, or by attempts to reintroduce a Heisenberg picture in spite of the Hamiltonian constraint.[27] The problem became pressing, though, in connection with the assumption of an ontic and kinematically complete wave function in quantum cosmology.[28]

The general wave functional $\Psi(h_{kl},matter)$ describes entanglement of geometry and matter. If we did have a succession of such quantum states (forming a quantum trajectory or quantum history), a very special, initially not entangled, state could explain an arrow of growing entanglement and decoherence – as usual. The resulting branching of the wave function according to an appropriate parameter *t* would then include branching states of spacetime geometry (that is, branching quasi-classical wave packets in the configuration space of three-geometries). Although there is no such time parameter any more, the metric $h_{kl}$ still contains a measure of metric time. Therefore, it describes a *physical* time dependence in the form of an entanglement of this measure with all other degrees of freedom – even for a formally time-less solution of (7).[29] For Friedmann universes, the expansion parameter *a*, which is part of the metric $h_{kl}$, is such an appropriate measure of time, but how does that help us to define an initial value problem for this static wave equation? The surprising answer is that this static equation is globally hyperbolic for Friedmann type universes on its infinite-dimensional gauge-free configuration space (which has therefore also been called "superspace") rather than on spacetime. The expansion parameter *a* or its logarithm appears as a time-like variable in this sense because of the unusual negative sign of its formal kinetic energy component.[30] Therefore, the Wheeler-DeWitt equation defines an "initial" value problem, for example at a small value of *a*. For a *modified* Wheeler-DeWitt equation, this possibility might even be ex-tended to *a = 0*. There is no conceptual difference between a (multiversal) Big Bang and a Big Crunch any more, since in the absence of a time parameter the wave function can only be a standing wave on configuration space (in spite of its intrinsic dynamics).

The metric tensor and other fields defined on a Friedmann sphere, *a = const,* may be represented by a four-dimensional multipole expansion, which is particularly useful for describing the very early, approximately homogeneous and isotropic universe.[31] In this case, one may conveniently model matter quantum mechanically by a massive scalar field *Φ($x_k$)*. The wave functional of the universe then assumes the form *Ψ(a,$Φ_0$,{$x_n$})*, where *$Φ_0$* is the homogeneous part of the scalar field, while *{$x_n$}* are all higher multipoles of geometry and matter. For the metric, only the tensor modes are geometrically meaningful, while the rest represents gauge degrees (here describing the propagation of spatial coordinates). The global hyperbolic nature with respect to all physical degrees of freedom becomes manifest in this representation.

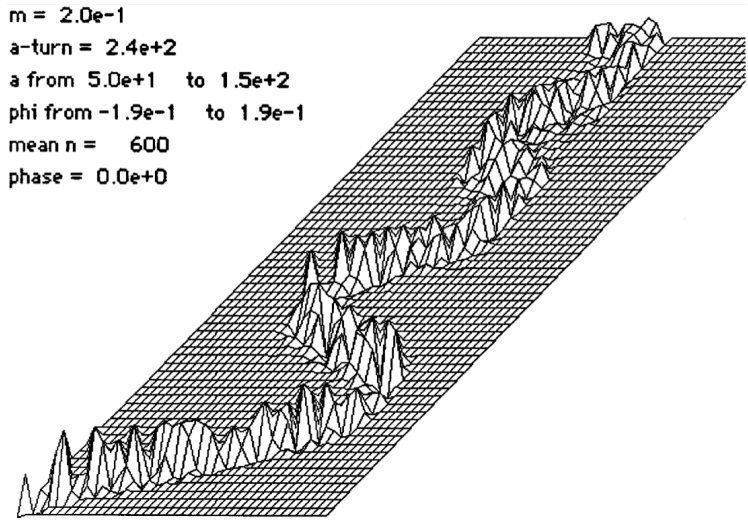

**Fig. 3:** Wave packet for a homogeneous massive scalar field amplitude *$Φ_0$* (plotted along the horizontal axis) dynamically evolving as a function of the time-like parameter *α = lna* that is part of the metric (second axis in this two-dimensional mini-superspace). The classical trajectory possesses a turning point above the plot region *50 ≤ a ≤ 150* – namely at about *a = 240* in this numerical example that represents an expanding and recontracting mini-universe. Wave mechanically, this corresponds to a reflection of the wave packet by a repulsive potential in (5) at this value of *a* (with the reflected wave being omitted in the plot). This reflection leads to considerable spreading of the "initial" wave packet. The causal order of these two legs of the trajectory is arbitrary, however, and the phase relations defining coherent wave packets could alternatively be chosen to give rise to a narrow wave packet for the second leg instead. Therefore, this (here not shown) formal spreading does *not* represent a physical arrow of time (From Ref. 1, Sect. 6.2.1.)

In a simple toy model one may neglect all higher multipoles in order to solve the Wheeler-DeWitt equation on the remaining two-dimensional "mini-superspace" formed by the two monopoles only. The remaining Hamiltonian represents an *a*-dependent harmonic oscillator for the variable $\Phi_0$, which allows one to construct adiabatically stable Gaussian wave packets ("coherent states").[32] Figure 3 depicts the propagation of such a wave packet with respect to the "time" variable $\alpha = \ln a$. This standing wave on mini-superspace mimics a timeless classical trajectory. However, the complete wave functional has to be expected to form a broad superposition of many such dynamically separated wave packets (a cosmologically early realization of "many worlds"). Note that these "worlds" are propagating wave packets rather than trajectories (as in DeWitt's or David Deutsch's understanding of "Many Worlds"). If the higher multipoles are also taken into account, the Wheeler-DeWitt equation may describe decoherence progressing with *a* – at first that of the monopole $\Phi_0$ and of *a* itself, although this approach requires effective renormalization procedures in this description.[33]

This "intrinsic dynamics" with respect to the time-like expansion parameter *a* has nothing as yet to do with the local dynamics in spacetime (controlled by proper times along time-like curves) that must be relevant for matter as soon as the metric becomes quasi-classical. In order to understand the relation between these two kinds of dynamics, one may apply a Born-Oppenheimer expansion in terms of the inverse Planck mass, which is large compared to all particle masses, in order to study the Wheeler-DeWitt wave function.[34] The Planck mass appears in the kinetic energy terms of all geometric degrees of freedom that appear in the Hamiltonian constraint. The formal expansion in terms of powers of $m_{Planck}^{-1/4}$ then defines an "adiabatic approximation" in analogy to the theory of molecular motion (with electron wave functions in the electrostatic fields of slowly moving nuclei). In most regions of configuration space (depending on the boundary conditions) one may further apply a WKB approximation to the "heavy" degrees of freedom *Q*. In this way one obtains an approximate solution of the type

$$\Psi(h_{kl}, matter) = \Psi(Q,q) = e^{iS(Q)}\chi(Q,q) \ , \tag{9}$$

where *S(Q)* is a solution of the Hamilton-Jacobi equations for *Q*. The remaining wave function $\chi(Q,q)$ depends only weakly on *Q*, while *q* describes all "light" (matter) varia-bles. Under these approximations one may derive from the Wheeler-DeWitt equation the adiabatic dependence of χ(Q,q) on Q in the form

$$i\nabla_Q S \cdot \nabla_Q \chi(Q,q) = h_Q \chi(Q,q) \quad . \tag{10}$$

The operator $h_Q$ is the weakly *Q*-dependent Hamiltonian for the matter variables *q*. This equation defines a new time parameter $t_{WKB}$ separately along all WKB trajectories (which define classical spacetimes) by the directional derivative

$$\frac{\partial}{\partial t_{WKB}} := \nabla_Q S \cdot \nabla_Q \quad . \tag{11}$$

In this way, one obtains from (10) a time-dependent global Schrödinger equation for matter with respect to the *derived* WKB time $t_{WKB}$.[26,28] This parameter defines a time co-ordinate in spacetime, since the classical trajectories *Q(t)* in the superspace of spatial geometries *Q* define spacetime geometries. Eq. (10) must also decribe the decoherence of superpositions of different WKB trajectories. Decoherence is also required to elimi-nate superpositions that are needed to define real waves function $e^{iS}\chi + e^{-iS}\chi^*$, which have to be expected from the real Wheeler-DeWitt equation under physically meaningful boundary conditions, in terms of the complex ones in (9).

In order to solve this derived time dependent Schrödinger equation along a given WKB trajectory, that is, in terms of a foliation of a classical spacetime that does in turn adia-batically depend on the evolving matter, one needs a (low entropy) initial condition in the region where the WKB approximation begins to apply. For this purpose, one would first have to solve the exact Wheeler-DeWitt equation (or its generalized version that may apply to some as yet elusive unified theory) as a function of *a* by using its funda-mental cosmic initial condition at *a = 0*. This might be done, for example, by using the multipole expansion on the Friedmann sphere, until one enters the WKB region (at some distance from *a = 0*), where this solution would provide initial conditions for the partial wave functions $\chi$ for all arising WKB trajectories. The derived time-dependent Schrö-dinger equation with respect to $t_{WKB}$ should then describe further decoherence of matter (the emergence of other quasi-classical properties), and thereby explain the origin of all other arrows of time. In particular, it must enforce decoherence of superpositions of any arising macroscopically different spacetimes, which would form separate quasi-classical "worlds".[26] It would also decohere conceivable *CPT* symmetric superpositions of black and white holes, which are analogous to parity eigenstates of chiral molecules, if these had ever come into existence.[16]

**Acknowledgement:** I wish to thank Claus Kiefer for his comments on an early draft of this manuscript, and Daniel Terno for a recent discussion.

**Note added after publication:** The "causal treatment" of black holes, used in Sect. 3 for an argument against the formation of event horizons and, therefore, the existence of an information loss paradox, has recently been corroborated by an explicit model that describes matter by means of a coherent scalar field.[35] An earlier attempt[36] had only modified my suggestion from the first arXiv versions of the present paper, whereas a similar scenario had already been proposed in 1976 (using a different model) by Ulrich Gerlach.[37] He assumed that the black hole finally settles down in a specific ground state that is not flat spacetime but would instead represent a "remnant". The essential assumption in all these models is the validity of relativistic causality even in the presence of retarded Hawking radiation. (In contrast to photon number eigenstates, general quantum field states possess a local basis that permits a definition of dynamical locality.[39]) Because of the simplicity of this argument, one may conjecture quite generally that event horizons can never form if matter is described by dynamically local QFT – in my opinion a very convenient and even plausible result. General relativity is thereby applied consistently by taking into account the energy flux represented by the Hawking radiation (understood as an expectation value if appropriate), while its here discussed consequence for the fate of black holes is then required by the dynamically arising light cone structure. However, some traditional concepts thereby turn out to be mere artifacts of classical GR: neither an event horizon at the Schwarzschild radius nor an internal black hole region ever forms if one takes into account the back reaction. Observers at fixed distances from the black hole would instead be affected by a heat bath of diverging temperature for *r→2M(t)*. Even though this heat bath may not be noticed by an inertial observer, the latter will be disrupted by the extreme tidal forces of the (from his point of view) rapidly shrinking black hole, and may later himself be transformed into Hawking radiation by some unitary mechanism close to *r=0* if Bekenstein and Hawking's prediction of thermal radiation remains valid. The phenomenon of black holes for external observers, on the other hand, is essentially unchanged, while there is no need for any quantum copying or "black hole complementarity" (which were sometimes proposed in order to resolve the "paradox").

The semi-classical description of black holes appears more realistic than a *quantum* gravitational collapse that neglects Hawking radiation, although this may also avoid a curvature singularity.[38] Both aspects may be relevant in the end, however.